\documentclass[12 pt]{article}

\begin{document}

{\centerline{\Large{\bf{Discovering Quantum Mechanics Once Again}}}}

\vspace{.20in}
{\centerline{\large{Ian Duck$\dagger$}}}
{\centerline{\it{Department of Physics and Astronomy}}}
{\centerline{\it{Rice University MS-61}}}
{\centerline{\it{Houston, Texas, USA, 77005-1892}}
\vspace{.15in}
{\centerline{July 15, 2003}}
\vspace{.25in}

\normalsize
\noindent
{\bf{Abstract:}}\\
We expand on a recent development 
by Hardy, in which quantum mechanics is derived from classical probability 
theory supplemented by a single new axiom, Hardy's Axiom 5.  Our scenario involves 
a `pretend world' with a `pretend' Heisenberg who seeks to construct a dynamical 
theory of probabilities and is lead -- seemingly inevitably -- to the 
Principles of Quantum Mechanics.

\vspace{.2in}
\noindent
{\bf{1}}{\hspace{.25in}}{\bf{Introduction}}

In two recent papers [1,2], Hardy shows how classical probability 
theory morphs directly into quantum theory, complete with full 
instructions for measurement and interpretation.  Hardy's demonstration 
employs five axioms.  The first four establish the basic 
classical probability theory in the vocabulary of a generalized 
Stern-Gerlach apparatus, which is the  paradigm of quantum mechanics.  Hardy 
introduces systems of $N$ states; each of dimension $K_N$ (=$N$ 
classically, =$N^2$ quantum mechanically); with subsystems $M\leq N$; 
and composite systems with $N=N_A N_B$ and dimension $K=K_A K_B$; and 
finally, Hardy introduces his crucial fifth axiom.  

Hardy's Axiom 5 requires that there 
exist a {\it{continuous}} reversible transformation between any two 
pure states.  Hardy emphasizes the marvelous purity of his derivation 
encompassed in the fact that the key word {\it{continuous}} is 
the sole genetic marker responsible for the profound distinction 
between classical probability theory and quantum mechanics.

Our purpose here -- following Hardy and essential earlier contributions 
by Caticha [3]-- is also 
to start with classical probability theory but to `derive' quantum mechanics in a less formal, 
austere and formidable way than Hardy.  We hope to `discover' in a `pretend world' 
a pathway equivalent to that of Hardy, which might appeal to the more heuristic and 
intuitive tastes of all those primarily interested in the physical as distinct from 
the more mathematical aspects of quantum mechanics. 

Of course, quantum mechanics must survive all these shenanigans completely 
unscathed.  What these efforts do hope to provide is not new physics, but 
perhaps a {\it{raison d'$\hat{e}$tre}}, an `understanding', of 
quantum mechanics -- an `interpretation' of quantum 
mechanics -- in line 
with the overwhelming, but not yet and probably never unanimous, consensus expressed by 
Fuchs and Peres [4] and many others, and still passionately debated [5]: 
specifically, that quantum mechanics is a theory of information propagation.  To say that 
quantum mechanics is `a' theory is to seriously understate 
the case.  Quantum mechanics will be seen to be 
`the' fundamental theory of information propagation.  The wave function is found to be the  
necessarily subjective encoding of information -- and when 
information changes, the wave function must be changed accordingly.  This 
interpretation puts an end at last to all quantum paradoxes.  They are now 
seen to be the result of a too literal -- even naive -- faith in our introduction to the 
subject via wave mechanical extensions of the classical objective world view.  

\vspace{.2in}
\noindent
{\bf{2\hspace{.25in}Quantum Mechanics from Information Theory}}

Here we present a `derivation' of quantum mechanics starting from 
information theory.  We find that it is surprisingly straightforward to reverse the roles of 
quantum mechanics and information.  The logic sequence is no longer\\
a) the discovery of the Heisenberg matrix mechanics underlying classical mechanics; and then of \\
b) Schroedinger wave mechanics; followed by \\ 
c) the verification of quantum mechanics in many stationary state situations; but\\
d) still plagued by paradoxical time-dependent situations; necessitating \\
e) a host of ad hoc `interpretations' of quantum mechanics to accommodate all paradoxes; and finally\\
f) resolution of all confusion by recognition of the role of the wave-function 
of quantum mechanics as the fundamental encoding and propagation of subjective information.

In fact the logic sequence can usefully be completely reversed.  We start, of course, with 
the essential advantage of knowing the desired   
results, and of having the necessary language and mathematical techniques 
already familiar from almost a century of development of quantum mechanics.  We are then able to `derive' 
quantum mechanics from the requirement that the information entropy of an isolated system 
should remain unchanged as it evolves in time.  We imagine a 
`pretend world' in which we `discover' quantum mechanics embedded in classical 
information probability.

We must be directly driven in this `pretend' world by a 
fundamental positivist philosophy to invent whatever tools 
are necessary to keep moving and -- hopefully -- to keep making progress.  These tools -- which are 
dramatically suggested, even demanded, by the formalism -- include unitary operators, hermitian 
generators, commutator brackets; all operating in a Hilbert space; then canonical 
commutation relations of $q$'s and their newly introduced $p$'s; then Heisenberg; then Schroedinger; 
then classical mechanics and eventually even the Lagrangian and the Principle of Least Action! 
Information theory explains the previously inexplicable:  Why a Principle of Least Action?      

The first step is to introduce the entropy of a macro-ensemble  
$$
S=-{\mbox{trace }}\{P {\mbox{ ln }}P\}
$$
where $P$ is the probability of a micro-ensemble.  None of these quantities will be fully 
specified but rather they will be 
explored.   $S$ is chosen to have the 
appropriate limiting values: $S=0$ if $P=1$ for the
simplest situation of a `pure state'.  Otherwise $S\geq 0$, and $S={\mbox{ ln }}N$ for a 
completely random mixture of $N$ such simplest situations each with probability $P=1/N$.  

The `trace' is a dimensional reduction by summing over all internal 
coordinates labeling the micro-ensembles which contribute to the macro-ensemble.  We 
are obviously cribbing the fundamental role 
of the density matrix and the implicit role of macro- and micro-ensembles.  

The next step is the stuff of legend: We 
imagine a brilliant 24 year old on a solitary vacation on a 
rain-swept rocky shore, daydreaming about the dynamical problem of 
calculating the probabilities $P$ rather than just assigning them, 
as we progress from thermodynamics to the next more fundamental theoretical 
level of statistical mechanics and ultimately to a fully dynamical theory.  Perhaps 
from analogy with Maxwell's dynamical 
equations for field amplitudes rather than for positive field energy 
densities, our young genius decides to:\\
a) factorize the probability $P$ into the suggestive form 
$$
P\Rightarrow\Psi^2=\Psi\times \Psi.
$$

This form has the virtue that $P\geq 0$ for arbitrary real $\Psi$.  Factorization 
leads to dynamical quantities satisfying the anticipated requirements of variational 
calculus where the virtual variations in $\Psi$ -- $\delta \Psi$ -- are unconstrained by 
the requirement that $P\geq 0$.  Here we have to borrow heavily from 
classical mechanics and the requirements on classical dynamical variables that they be 
continuous and differentiable.  This motivates us to search for an analytic dynamical theory 
for the probabilities $P$, but that is impossible.  The normalization of the 
probabilities -- ${\mbox{tr}}P=1$ -- is a {\it{holonomic}} 
constraint (i.e., an {\it{equality}} constraining the candidate generalized coordinates $P$) 
in Goldstein's classification [6], and is manageable 
in the context of classical mechanics simply by eliminating one $P$.  However, the positivity 
condition on the probabilities -- $P\geq 0$ -- as an {\it{inequality}} is a 
{\it{nonholonomic}} constraint which ``there is no 
general way of attacking''.

A factorizable probability is 
the especially simple 
case of a pure state but it is 
an immensely instructive warm-up exercise.  Proceeding, \\
b) we require $\Psi$ to be more 
than just $\sqrt{P}$.  We require $\Psi$ to be analytic and differentiable in 
its variables, and therefore to have the possibility of sign-changes 
where $P=0$.\\
c) The invariance of $P$ and $S$ under a simple sign-change of $\Psi$ is 
further extended by allowing $\Psi$ to be complex so $P\rightarrow \Psi\Psi^{*}$.  
Why?  Why complex numbers?  Our basic response is:  Why not?  Our only requirement was 
that $P\geq 0$, so to choose $\Psi$ real was to overconstrain it.  With 
complex $\Psi$ \\
d) the full invariance group of $P$ and $S$ is the group of all unitary 
transformations $U$ on $\Psi$, with $U^{\dagger}U=1$
$$
\Psi\rightarrow\Psi'=U\Psi,
$$
and the merit of allowing $\Psi$ to be complex becomes evident.

Without complex 
$\Psi$, the invariance group of $P$ and $S$ is just the trivial change of sign 
$\Psi\rightarrow -\Psi$, and our search for a dynamical theory of $P$ 
comes to a grinding halt.  With complex $\Psi$ we are on familiar ground.  
$\Psi$ is the state function for the 
macro-ensemble of the density matrix, here reduced (trivialized) to 
the simplest pure-state micro-ensemble.  We will relax these 
restrictions soon.\\
e) For the system to evolve smoothly in time without change of entropy of information, 
requires the time evolution of the state function $\Psi$ to be a unitary transformation
$$
|\Psi(t=0)>\Rightarrow |\Psi(t)>=U(t)|\Psi(0)>\equiv e^{-i Ht}|\Psi(0)>  
$$
(going over to the familiar Dirac notation).  The invariance of the entropy 
is made clear in this way for our simplest system of a single micro-ensemble, and 
is maintained for a mixture of noninteracting micro-ensembles.  Caticha [see 3] enforces 
this requirement by requiring the Hilbert space as the ``uniquely natural'' choice.  
Caticha formalizes this development, but we proceed with our `pretend' 
discovery scenario.

  Here we have introduced the hermitian 
generator of infinitesimal time translations $H= H^\dagger$.
This is by definition the Hamiltonian, but we have not imposed anything 
except hermiticity in the new information based dynamics.  The familiar Hamiltonian 
dynamics will be a product 
of the information theory.  We have also kept to the simplest possible case by 
taking $H$ to be time independent. 

Next,\\
f) the Heisenberg equations of motion follow immediately from
\begin{eqnarray*}
<x(t)> &=& \mbox{ tr }\{x(0)P(t)\}\\
       &=& \mbox{ tr }\{x(0) U|\Psi(0)><\Psi(0)|U^{\dagger}\}\\
       &=& \mbox{ tr }\{U^{\dagger}x(0)U |\Psi(0)><\Psi(0)|\},
\end{eqnarray*}
so
$$
x(t)=U^{\dagger}x(0)U
$$
and 
\begin{eqnarray*}
\dot{x}(t)&=&iU^{\dagger}[H,x(0)]_{CB} U\\
          &=&i[H,x(t)]_{CB}.       
\end{eqnarray*}
This is the familiar Heisenberg matrix equation of motion involving the commutator bracket of 
$x(t)$ with the generator of time translations $H$.
  
Proceeding {\it{ab initio}} from 
a Principle of Stationary Information Entropy seems logically possible, but we forgo the exercise by   
identifying $H$ with a non-trivial Hamiltonian and introducing for each 
generalized coordinate $x$ the canonically conjugate 
operator $p$ which diagonalizes the commutation relations; then we get the structure of Heisenberg 
matrix mechanics; and from that the Schroedinger derivative representation of the commutation relations, 
and wave mechanics; and finally the classical limit of commutator brackets as Poisson brackets.  So 
we could also even `discover' classical mechanics in this inverted logic.  
  
The Hamiltonian is paramount in this formulation, as it is in ordinary non-relativistic 
quantum mechanics.  It remains to find the Lagrangian and the Action Principle, 
both of which seem somewhat contrived and {\it{ad hoc}} in the usual derivation of 
non-relativistic quantum mechanics.  Helmholtz and Gibbs [7]  
have showed that the stationary minimum of 
the free energy A=E-TS+PV [8] fulfills the role of Lagrangian in 
reversible chemical reactions.  Extending the invariance group of the entropy to the Lorentz group 
will be useful also.

What can be said when the probability $P$ is not simply factorizable?  The density matrix 
[9] must then be written as a sum over micro-ensembles $\Psi_j$ each weighted with its own 
positive probability $P_j$.  (Keep in mind that the $P_j$ are now real numbers satisfying $P_j\geq 0$ 
and $\sum_j P_j =1$.)  
$$
\rho(t=0)=\sum_{j}\{|\Psi_j>P_j<\Psi_j|\} 
$$
in the usual notation.  All the above equations are {\it{required}} 
to survive as
\begin{eqnarray*}
<x(t)>&\equiv& \mbox{ tr }\{U^{\dagger}x(0)U\rho(0)\}\\
<\dot{x}(t)>&=& i\mbox{ tr } \{U^{\dagger}[H,x(0)]_{CB}U \rho(0)\}\\      
            &=& i\mbox{ tr } \{[H,x(t)]_{CB} \rho(0)\}.
\end{eqnarray*}
  
The conditions for this survival are profound:  we require a Hilbert space of the  
$\Psi$'s and the corresponding operator or matrix representation of the $U$'s.  We 
introduced the notation somewhat gratuitously in the above `pure'-case, but now the 
necessity of the full Hilbert space formalism is clear.  If we were to make any 
progress at all in our `pretend world' in the non-trivial `mixed'-case we 
would have had to invent Hilbert space and operator $U$'s at this point.

What is a Hilbert space?  It is the complex vector 
space of the eigenvectors $|\Psi_j>$ of 
the hermitian operator $H$, so 
$$
H|\Psi_j>=E_j|\Psi_j>,
$$
which have the quadratic norm
$$
<\Psi_j|\Psi_k> =\delta(j,k)
$$ 
and so are orthogonal.  They are complete
$$
\sum_j |\Psi_j><\Psi_j| ={\bf{1}},
$$
where {\bf{1}} must be `interpreted'.  We are confronted by the seemingly 
absurd, but in fact profound, Clintonesque question:  What {\it{is}} {\bf{1}}?   

Two simplest examples are useful to keep in mind: \\
a) The simplest is the 
two dimensional space of a spin-$\frac{1}{2}$ particle with\\
\begin{eqnarray*}
S_x&=&\frac{\sigma_x}{2}, \mbox{ etc. for } y,z;\\
H&=&{\Delta} S_z  
\end{eqnarray*}
with two eigenstates 
$$
|\Psi_{1}>\equiv \alpha=\left( \begin{array}{c} 1\\0\end{array}\right) \mbox{ and } 
|\Psi_{2}>\equiv \beta =\left( \begin{array}{c} 0\\1\end{array}\right).
$$
Orthonormalization and completeness are apparent.  Complex representations 
could equally well have been chosen, e.g., the eigenstates of 
$\sigma_x$ or $\sigma_y$ instead of those for $\sigma_z$.  The 
complex vector space in this simplest case 
is spanned by two 2-dimensional orthonormal and complete basis-vectors, such as 
$\alpha$ and $\beta$.\\
\noindent
b) A less trivial example is the free particle in three unconstrained dimensions.  
The Hamiltonian is 
$$
H=\frac{p^2}{2m}
$$
and we choose simultaneous energy and momentum eigenstates
$$
|\Psi_{\vec p}(\vec x)>=e^{i{\vec p}\cdot{\vec x}}.
$$  
These have continuum orthonormalization 
\begin{eqnarray*}
\sum_ {\vec x}<\Psi_{\vec p}(\vec x)|\Psi_{\vec k}(\vec x)>&=&\int d^3 x e^{i({\vec k}-{\vec p})\cdot{\vec x}}\\
                                                            &=&(2\pi)^3\delta^3({\vec k}-{\vec p}),
\end{eqnarray*}
and the very similar completeness relation 
\begin{eqnarray*}
\int \frac{d^3 p}{(2\pi)^3} e^{i{\vec p}\cdot({\vec x}-{\vec y})}=\delta^3({\vec x}-{\vec y}).
\end{eqnarray*}
In this case, the complex vector space is spanned by a mind-boggling array of 
basis-vectors: there is a triple-infinity of orthonormal 
and complete basis-vectors labeled by ${\vec p}$, each with a triple-infinity of 
components labeled by ${\vec x}$ (or vice versa).  

We conclude that a Principle of Stationary Information
Entropy, coupled with some inspired (but {\it{a posteriori}} inevitable) 
requirements during the analysis, 
can completely reverse the logical structure of quantum mechanics.  
With this approach, there is no `interpretation' of quantum mechanics; nor 
need there be any hesitation or delay in this `pretend world'
of making all the usual applications.  A Hilbert space is required.  Heisenberg 
matrix-commutator mechanics is required.  We can imagine that the concept of generalized
coordinates and their conjugate momenta would naturally occur, even 
without classical mechanics, as an optimal minimum set of
non-commuting variables defined to diagonalize the commutation relations in a 
standard way.  Schroedinger's differential representation 
would be next and with it, all of wave mechanics and its intuitive (but 
occasionally misleading) guidance.  And -- as now -- we can imagine a compelling route
even to classical physics, but with a deeper understanding of the 
Principle of Least Action.

\vspace{.2in}
\noindent
{\bf{3\hspace{.25in}Adding Structure}}

Now we have to ask:  What is $\Psi_j$? and what is $\sum_j$ ? In fact, 
what is $j$?

Our answers to these questions must satisfy the requirement that ``$\Psi$ is the 
complex resolution of unity'' so we have completeness 
$$
{\bf{1}}=\sum_j |\Psi_j><\Psi_j|, 
$$
and orthogonality 
$$
<\Psi_j|\Psi_k>=\delta(j,k).
$$
The density matrix is constructed as   
$$
{\bf{\rho}}=\sum_j|\Psi_j>P_j<\Psi_j|
$$
with $P_j\geq 0$ and $\sum_j P_j =1$.
Our target structure will be ordinary quantum mechanics which will 
make its appearance in the old-fashioned but explicit and intuitive 
Fock-representation.

The simplest case has dimension $D=2$ and $\Psi_{j}$'s which are 
two states specified by the 
elementary Pauli spinors.  It is a simple but instructive example and 
proves to be a faithful guide to any level of complexity.  We have 
the standard representation  
$$
|\Psi_{1}> \equiv \alpha=\left( \begin{array}{c}1\\
                                    0\end{array} \right)
\mbox{ and }|\Psi_{2}> \equiv \beta =\left( \begin{array}{c}0\\
                                    1\end{array} \right).
$$ 
The two dimensional (real-)resolution of unity is 
\begin{eqnarray*}
{\bf{1}}&=&|\Psi_{1}> <\Psi_{1}|+|\Psi_{2}><\Psi_{2}|\\
&=&\left( \begin{array}{cc}1&0\\0&1\end{array} \right),
\end{eqnarray*}
and the density matrix is
\begin{eqnarray*}
{\bf{\rho}}&=& |\Psi_{1}>P_{1}<\Psi_{1}|+|\Psi_{2}>P_{2}<\Psi_{2}|\\
                                 &=&    \left( \begin{array}{cc}P_{1}&0\\0&P_{2}\end{array} \right).
\end{eqnarray*}
The unitary time evolution of these states which leaves $\rho$ unchanged is 
generated by the no-interaction Hamiltonian 
$$
H_{0} =\frac{\Delta}{2}\sigma_{z}
                           ={\frac{\Delta}{2}}\left( \begin{array}{cc}1&0\\0&-1\end{array} \right).
$$

An interaction between the two states would involve $\sigma_x$ and/or $\sigma_y$ 
and would induce continuous transformations between the two states, which were 
chosen to diagonalize the initial density matrix.   Hardy [1,2] identifies the existence of these 
{\it{continuous}} transformations between basis states as the essential 
distinction between classical probability theory and quantum mechanics.  It 
is seen to arise here from the extension of the `resolutions of unity' -- suggested by the requirement 
that the sought-for conjectured classical dynamical variables ($\sim \sqrt{P}$) satisfy only holonomic 
constraints -- to the `complex resolutions of unity'.  These must be included  
for the sake of logical completeness.   

We can extend this simplest example in a multitude of  directions.\\
a)  The first extension is 
almost trivial: generalize the 2-D $SU(2)$ 
example above to other similar groups including the familiar $O(3)$ and $SU(3)$.\\  
b) A second extension is to the 
interesting dynamical problems which arise when the density matrix of two (or more) {\it{a priori}} 
independent systems is defined in the direct product space of the two systems as
$$
{\bf{\rho}}(1,2)={\bf{\rho}}(1)\bigotimes {\bf{\rho}}(2).
$$
This requires basis representations 
$$
\Psi(1,2)=\Psi(1)\bigotimes \Psi(2)
$$
and the Hamiltonian
$$
H(1,2)=H_{0}(1)\bigoplus H_{0}(2)\bigoplus H_{int}(1,2)
$$
which dynamically couples the two systems.  Such dual -- but isolated -- systems 
still fall short of a model for the full measurement process [10].  We should not be 
surprised or disappointed.  Standard quantum mechanics is our limited goal.

A third extension required for continuous variables \\
c) like $\vec{x}$ and 
$\vec{p}$ is somewhat different but is directly suggested by the $SU(2)$ example: it involves
judicious replacements of sums by integrals and Kronecker $\delta_{K}(j,j')$ 
by Dirac  $\delta_{D}(p-p')$.  For example, the `complex resolution of unity' for a 
probability distribution defined in momentum space now becomes 
$$
{\bf{1}}\Rightarrow (2\pi)^3 \delta^3(p-k)=\int d^{3}x e^{i(\vec{p}-\vec{k})\cdot \vec{x}}
$$
and 
$$
\Psi_{j}(\vec{p})\Rightarrow e^{i\vec{p}\cdot \vec{x}}.
$$

One might well ask: Where did the plane wave solutions come from in this `pretend world'?  
Of course, we have the answer from quantum mechanics as we have been taught it; but how would we 
arrive at this result starting with a stationary 
probability, factored into `complex resolutions of unity'?  

We must start with a  
dimensionality chosen for the problem at hand.  This is where physics, judgement, and 
discovery enters.  The next step is to require the time evolution operator to be unitary, and 
the generator to be a hermitian Hamiltonian operator whose commutator with $x$ is the velocity $v$.  
We can follow two paths here:  The easiest one is to take classical mechanics as 
already known, and simply use the Hamiltonian $H=p^2/2m$.  Then the 
commutation relation requires $p=\hbar \nabla/i$, and the choice of plane waves as `resolvent' 
functions is dictated by the resulting Hilbert space.  Again there are judgements 
to be made, and we choose a rectangular basis which diagonalizes the linear momenta.  This choice 
replaces all operators by their eigenvalues.  

A second path suggests itself, as mentioned above, of deducing classical 
mechanics {\it{ab initio}} from the factorization requirement, and diagonalizing the 
commutation relations.  In this way, we are lead to `discover' the generalized momentum $p_x$ conjugate 
to the generalized coordinate $x$.  We are limited in this way to a unit metric 
in generalizing sums to integrals, but we do get a toehold and could subsequently change 
to a basis of angular momentum eigenstates for example.       

Traditional quantum calculations are quite conveniently made in this 
density function representation.  We give a  brief heuristic sketch of transitions 
between basis states induced by a non-diagonal interaction Hamiltonian $H'$.  The 
initial density matrix evolves in time to  
\begin{eqnarray*}
\rho(0)&\equiv&|\Psi_{j}><\Psi_{j}|\\
&\rightarrow& \rho(t)=e^{-iH't}\rho(0)e^{+iH't}\\
&\rightarrow& \sum_{m}|\Psi_{m}>P_{m}(t)<\Psi_{m}|.
\end{eqnarray*}
The probability $P_{k}(t)$ at time $t$ of a state $k\neq j$ is 
\begin{eqnarray*}
P_{k}(t)&=& tr |\Psi_{k}><\Psi_{k}|\rho(t)\\
        &=&|<\Psi_{k}|e^{-iH't}|\Psi_{j}>|^2\\
        &\simeq& t^2|<\Psi_{k}|H'|\Psi_{j}>|^2,
\end{eqnarray*}
proportional in first order to the absolute square of the matrix 
element of the perturbing interaction Hamiltonian.  The factor 
$t^2$ requires some care in application to realistic energy conserving 
transitions but we won't pursue that.  We should perhaps better recover 
the Schroedinger equation directly to evaluate 
$$
e^{-iH't}|\Psi_{j}>
$$
in terms of the energy eigenstates of the full Hamiltonian.  In either case, we 
return directly to the usual quantum results.   

\vspace{.20in}
\noindent
{\bf{4\hspace{.25in}Concluding Remarks}}

Let us summarize and reiterate what has been done.  We start with the 
information entropy of a macro-ensemble
$$
S=-\sum_{j} \{P_{j}{\mbox{ ln }}P_{j}\} \geq 0,
$$
where $j$ enumerates the micro-ensembles occurring in the macro-ensemble 
with probability $P_{j}$.  $P_{j}$ must satisfy the obvious restrictions
$$
0\leq P_{j}\leq 1 \mbox{  and  } \sum_{j} P_{j}=1.
$$
We then embark on a `pretend'-journey of `discovery' resulting in quantum 
mechanics.

Our goal is to find a dynamical theory governing the system.  We 
rule out the probabilities $P_{j}$ themselves as candidate fundamental 
variables of such a dynamical theory.  The reason is familiar from classical mechanics: 
the inequality $P_{j}\geq 0$ is a {\it{non-holonomic}} constraint
and  ``there is no general way'' of attacking such problems.  This leads us to factorize the 
probabilities ultimately to 
$$
P_{j}\rightarrow \Psi_{j}^2 \mbox{ and further to  }\rightarrow \Psi_{j}\Psi_{j}^*.
$$
\\
a) Why factorization?  To satisfy the non-holonomic constraint identically.\\
\noindent
b) Why complex factors?  This generalization is obviously permissible and thus a logical necessity. \\ 
\noindent 
c) What factors?  The answer to this question builds upon experience with the simplest system 
imaginable: the 2-dimensional complex space of spin-$\frac{1}{2}$.   Finally we can conclude that the 
$\Psi_{j}$'s are a complete orthonormal set of complex basis vectors in the Hilbert space generated 
by the Hamiltonian describing the dynamics of the system under consideration. \\
\noindent
d) What dynamics?  To preserve the entropy of a macro-ensemble, we require the time dependence of the $\Psi_{j}$
to be a unitary transformation generated by the hermitian time evolution operator 
(i.e., the Hamiltonian) of the system.  The actual {\it{choice}} of the Hamiltonian is not made by the 
quantum theory {\it{per se}} but becomes an act of creative judgement subject only 
to the achievement of interesting results.  \\
\noindent
e) Elementary considerations lead directly to the Heisenberg equations of motion involving the commutator 
with the Hamiltonian.\\
\noindent    
f) The $\Psi_{j}(t)$ satisfy the Schroedinger equation and in the usual way constitute the required 
Hilbert space vectors.\\

We could continue in this way, or we could return to ordinary quantum mechanics with 
the -- perhaps not so new -- understanding of the wave functions as `complex resolutions of unity'
-- i.e., projection operators --  onto each particular micro-ensemble in the macro-ensemble 
under consideration.

What has been gained is not a new quantum mechanics, but a reason for the existence of the 
old one.  The existence of quantum mechanics 
is necessary in order for there to be a fundamental dynamical theory governing 
the elementary probabilities in the Shannon Information Entropy.  In addition, this 
derivation justifies the subjective interpretation of the wave function as the 
encoding of information.  Even further, such fundamental entities as the Heisenberg 
matrix equations of motion appear naturally, suggesting the fundamental commutation 
relations diagonalized by the introduction of the momentum canonically 
conjugate to each independent coordinate;  and 
even the derivation of classical mechanics from this quantum mechanics, 
and an alternative to the Principle of Least Action; all from a Principle of 
Stationary Entropy.

As advertized, our development is intended to be a `pretend' voyage of discovery, 
not a mathematical derivation of quantum mechanics.  Caticha [3] points out the 
many logical shortcomings in our too-facile assumptions, 
some of which we describe below.  First, however, let us point out that 
the possibilities employed in our 
scenario do exist as a path through the maze.  Our specific {\it{choice}}  
of complex numbers, Hilbert spaces, and unitary transformations turns out to be 
{\it{sufficient}} and {\it{self-consistent}} but not proven to be 
necessary.  Caticha [3] points out the possibility of Clifford algebras [11] 
of real vectors {\it{inter alia}}.  These do have a possible presence in extensions 
of quantum mechanics to include Weyl spinors and Grassmann variables, but 
the onus so far has been on the conformability of these structures with the 
pre-existing structure of quantum mechanics.

Our philosophy is a pragmatic positivist one.  In this 
view, every exception is to be viewed not as a barrier to progress, but as 
an opportunity.  It is an appeal to a Correspondence Principle.  
At the same time, we have to acknowledge the possibility that some arcana 
really are really mysterious.    

\vspace{.20in}
\noindent
{\bf{5}}\hspace{.25in}{\bf{Acknowledgements}}

I am deeply indebted to Lucien Hardy's monumental tome [1] which 
I now feel more fully able to appreciate.  The earlier fundamental 
work by Ariel Caticha [3] was  particularly valuable, as was the popular 
account in PHYSICS TODAY by Chris Fuchs and Asher Peres [4].  In addition, 
Asher Peres has kindly brought to my attention his similar scenario [12] 
in which probabilities $P_{mn}$ are shown to be absolute squares of 
unitary matrices $U_{mn}$, which are defined by generalized Stern-Gerlach 
devices.   
 
\vspace{.20in}
\noindent
{\bf{References:}}

\noindent
$\dagger$) duck@rice.edu\\
1) L. Hardy, {\it{Quantum Theory From Five Easy Axioms}}, quant-ph/0101012.  
The first four axioms, in brief: (1) define {\it{probability}} as the relative frequency of occurence
in an ensemble; (2) require {\it{simplicity}} in terms of minimum degrees of 
freedom K for given dimensionality N; (3) have {\it{subsystems}} obeying the same rules; so (4) 
{\it{composite systems}} can be directly composed of subsystems.\\
2) L. Hardy, {\it{Why Quantum Theory?}}, quant-ph/0111068.\\
3) A. Caticha, {\it{Probability and entropy in quantum theory}}, quant-ph/9808023.\\
4) C. A. Fuchs and A. Peres, PHYSICS TODAY, March (2000).  For immediate responses, see 
PHYSICS TODAY, September (2000).\\
5) M. Gell-Mann and J. B. Hartle, Phys. Rev. D{\bf{47}}, 3345 (1993); also: 
R. Omnes, {\it{Understanding Quantum Mechanics}}, Princeton U. P., Princeton,
NJ (1999); S. Goldstein, PHYSICS TODAY, March and April (1998) and responses
PHYSICS TODAY, February (1999).\\
6) H. Goldstein, {\it{Classical Mechanics}}, Addison-Wesley, Reading, MA (1981).\\
7) W. Yourgrau and S. Mandelstam, {\it{Variational Principles in Dynamics  and 
Quantum Theory}}, Pitman, London (1960), p.173, 176.\\
8) K. Huang, {\it{Statistical Mechanics}}, Wiley, NY (1987).\\
9) R. P. Feynman, {\it{Statistical Mechanics}}, Benjamin, Reading, MA (1972), Ch.2.\\ 
10) B. Misra, I. Prigogine, and M. Courbage, {\it{Lyapounov Variable:  Entropy and 
Measurement in Quantum Mechanics}}, Proc. Nat. Acad. of Sciences USA, {\bf{76}}, 
4768 (1979); reprinted in: J. A. Wheeler and W. H. Zurek, {\it{Quantum Theory and 
Measurement}}, Princeton U. P., Princeton, NJ (1983), p.687.\\
11) N. Prakash, {\it{Mathematical Perspectives on Theoretical Physics}}, 
Tata McGraw-Hill, New Delhi (2000), p.157, 288.\\
12) A. Peres, Nuclear Physics B (Proc. Suppl.){\bf{6}}, 243-245 (1989).

\end{document}